\documentclass[runningheads]{llncs}

\usepackage{graphicx}
\usepackage{array}
\usepackage{diagbox}
\usepackage{here}
\usepackage{indentfirst}
\usepackage{url}

\begin{document}
%
\title{SpatialViewer: A Remote Work Sharing Tool that Considers Intimacy Among Workers\thanks{Supported by Seiwa Business Co., Ltd. and Workscape Lab.}}
\titlerunning{SpatialViewer}
%
%
\author{Sicheng Li \and
Yudai Makioka \and
Kyousuke Kobayashi \and
Haoran Xie \and
Kentaro Takashima}
\authorrunning{S. Li et al.}
%
\institute{Japan Advanced Institute of Science and
Technology, Ishikawa, Japan
\email{ktaka@jaist.ac.jp}}

\maketitle              
\begin{abstract}
Due to the influence of the new coronavirus disease (COVID-19), teleworking has been expanding rapidly. Although existing interactive remote working systems are convenient, they do not allow users to adjust their spatial distance to team members at will, 
 and they ignore the discomfort caused by different levels of intimacy. To solve this issue, we propose a telework support system using spatial augmented reality technology. This system calibrates the space in which videos are projected with real space and adjusts the spatial distance between users by changing the position of projections. Users can switch the projection position of the video using hand-wave gestures. We also synchronize audio according to distance to further emphasize the sense of space within the remote interaction: the distance between projection position and user is inversely proportional to the audio volume. We conducted a telework experiment and a questionnaire survey to evaluate our system. The results show that the system enables users to adjust distance according to intimacy and thus improve the users' comfort\footnote{This is a pre-peer review, pre-print version of this article. The final authenticated version is available online at: \url{https://doi.org/10.1007/978-3-030-77599-5_5}}.

\end{abstract}

\section{Introduction}
Currently, new coronavirus disease (COVID-19) infections are exploding. To prevent and control infections, the normal offline working environment has been inhibited in a growing number of companies. Teleworking from home is a burgeoning way of working in many industries. Teleworking has the advantages of creating a comfortable working environment and reducing interruption and distractibility, while having the disadvantages of impeding communication with colleagues and causing a sense of loneliness. Various applications have been proposed to encourage communication in teleworking, including commercial applications such as Zoom\cite{Zoom} and Cisco Webex\cite{CiscoWebex} for remote meetings and Slack\cite{Slack} and Microsoft Teams\cite{MicrosoftTeams} for collaboration and business contact. These applications satisfy the formal communication needs of the business operations, which is the primary demand for teleworking, but not the only demand.

It is still challenging to realize a teleworking environment that resembles a real office using these applications. It is difficult to remain aware of each other's state, listen to surrounding sounds, or occasionally consult with colleagues. In general, users launch these applications only when they need to hold a meeting or business conversation. Existing remote video-conferencing tools can certainly be used to remain aware of team members' current situations 
 if they are used continually. However, these applications do not consider the different levels of intimacy among team members. These applications do not allow remote workers to adjust the strength of their awareness of different partners.
 
 When remote workers connect to these regular teleworking support platforms to share their daily situations, all workers on a large team or in a department are displayed on one screen and mute themselves on group calls until they need to speak. Workers might feel oppressed and become tired when facing team members with whom they are unfamiliar. Workers sometimes want to share news only with more intimate colleagues and may be reluctant to share with others. In real life, people have a sense of interpersonal distance. In an office environment, personal space is secured by adjusting the distance between people according to their relationships. If remote workers could make such adjustments virtually, they would be able to remain more comfortable while teleworking. 

\begin{figure}[h]
\centering
\includegraphics[width=0.8\textwidth]{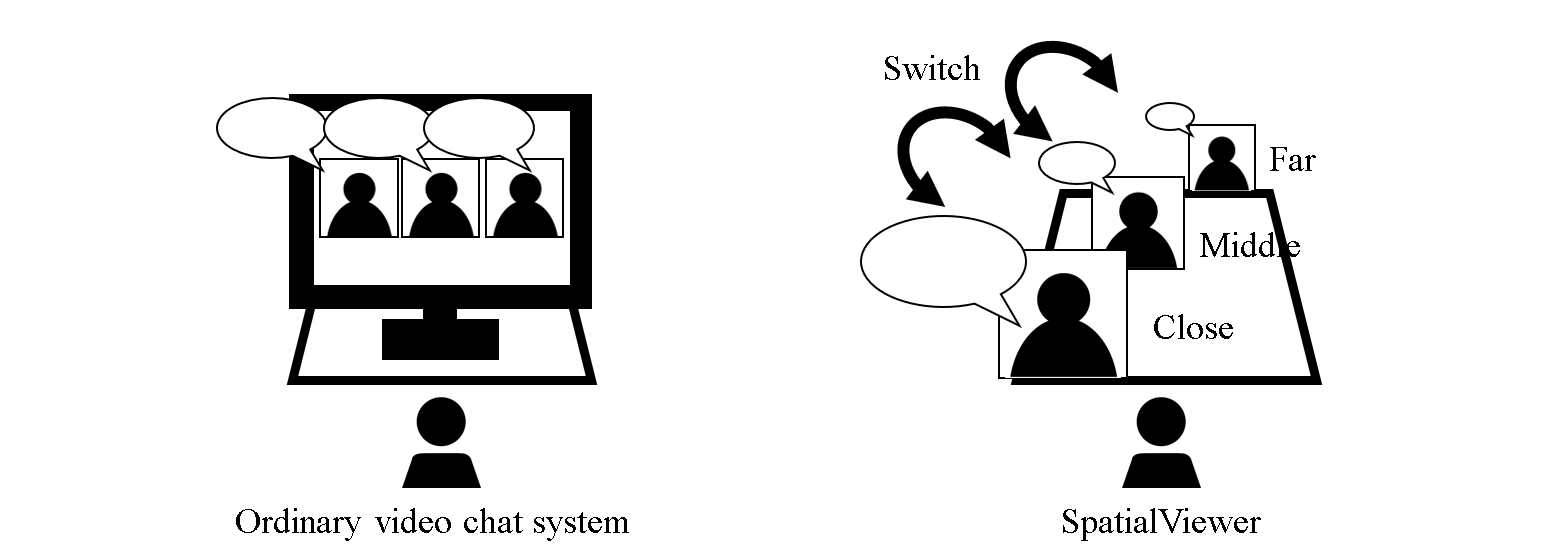}
\caption{Our proposal} \label{fig1}
\end{figure}

This study proposes a novel telework support system called SpatialViewer that considers the different levels of intimacy among users. SpatialViewer allows remote workers to adjust interpersonal distances and audio volumes during mutual video connections (See Fig.~\ref{fig1}). The system achieves video and audio spatialization by integrating with a prototype of a video chat system. 
 The proposed system projects videos of colleagues onto items in the user's actual desk environment using Spatial Augmented Reality(SAR). SpatialViewer enables users to adjust the distance of the projected items from themselves and to continue teleworking comfortably for a long time. Besides, it allows users focus on the colleagues who users needed to share information and support smooth communication.

The main contributions of the proposed SpatialViewer system are as follows:
\begin{itemize}
\item We pay attention to the influence of interpersonal distances among team members and try to replicate these interpersonal distances in remote interactions.
\item We propose a novel teleworking system that allows users to arrange the video and audio of team members according to the desired spatial intimacy.
\item We develop a user interface that supports gesture control to switch the location of each video flexibly.
\end{itemize}

\section{Related Works}
In recent years, SAR technology has been widely used in the field of human-computer interaction (HCI). It realizes to transfer a virtual user interface into real space. 
 Many studies have focused on developing HCI based on Augmented Reality(AR) 
\cite{9122361}\cite{9122335}\cite{9122337}. Other research has worked on situation sharing and communication support in a distributed environment, using SAR and virtual reality (VR) technology\cite{KanaMisawa2013}\cite{Murata2018} and to visualize the working environment constantly, thus facilitating remote communication \cite{morrison2020}.

For example, Pesja et al.\cite{Pesja2016} proposed a telepresence system that makes remote conversation feel more immediate by projecting the image of the interlocutor in real space. In their research, 
 the experience of a face-to-face conversation is realized by projecting a full-scale image of a remote interlocutor onto an assigned location. Yao et al.\cite {Yao2018} proposed a system that supports coordination among coworkers for collaborative teleworking and improves educational experiences by visualizing the sight lines of collaborative users. Hoppe et al.\cite{Hoppe2021} proposed a collaborative system that enables multiple users to interact face-to-face simultaneously by moving users' avatars between different locations, so they can each share their perspective.

In addition, some works have visualized remote communicators by reproducing them as avatars. For example, some works proposed methods by which users in remote locations can perceive themselves to be sharing space using VR, mixed reality (MR) devices, and robots 
 \cite{Murata2018}\cite{Jones2020}. Misawa et al.\cite{KanaMisawa2013} proposed a telepresence method using a robot equipped with a simulated human face to act as the avatar of remote users in support of nonverbal communication. In addition, Piumsomboon et al.\cite{Piumsomboon2018} developed a remote collaboration system that visualizes the gaze direction and body gestures of a remote user wearing a VR headset; using avatar representation, the size and orientation of the remote user avatar are changed 
 to stay in the AR user's field of sight. Ruvimova et al. proposed a platform that aims to improve business performance by building a virtual environment away from the office using virtual reality (VR)\cite{Ruvimova2020}.

However, it was a challenging for previous studies to adjust the degree of the information sharing and intimacy among team members. In the physical environment, the degree of presence and the amount of information shared with other people are regulated mainly by interpersonal distance. Interpersonal distances include close distances (0-45 cm) for close relationships, individual distances (45-120 cm) for private communication, social distances (1.2-3.6 m) for formal information sharing, and public distances (more than 3.6 m) where no personal interaction takes place\cite{doi:10.1177/001316446602600462}. As interpersonal distance decreases, the other person’s presence and influence 
 increase. With a close interpersonal distance to other users, one’s situation is shared and the probability of communication increases. It is important for users to adjust their interpersonal distance with coworkers according to their relationship and communication situation. Even within a virtual space, people have been reported to perform adaptive behaviors to maintain an individual distance when interacting with agents in the virtual space on the screen\cite{Rapuano2021}.

\section{Proposed System}
\subsection{System Overview}
\begin{figure}[H]
 \centering
\includegraphics[width=\textwidth]{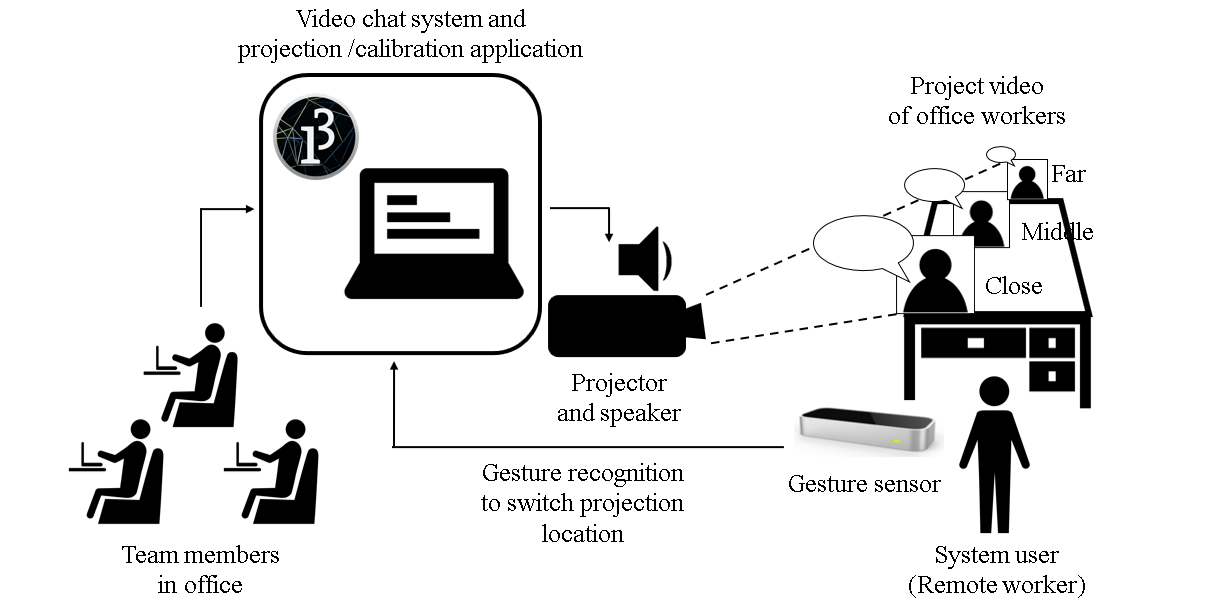}
\caption{Overview of the proposed SpatialViewer system} \label{fig2}
\end{figure}

In this research, we propose a novel teleworking system that allows users to adjust their distance to each team member according to their interpersonal relationships. The system overview is shown in in Fig.~\ref{fig2}. The system projects the video images of remote team members onto pre-set items placed in the teleworkers' desk environments. Users can customize the location of each item and place it nearby (i.e. close to the display of the working PC) or far away (i.e. on a partition above the desk that is out of sight), depending on their intimacy and the need for information sharing with each team member. Users can also switch the location of team members' video images while the system is in use by gesture control. The volume of each team member's audio is changed automatically according to the projection distance.

\subsection{System Functions}
\subsubsection{Video and Audio Connection}
\begin{figure}[h]
\centering
\includegraphics[width=0.5\textwidth]{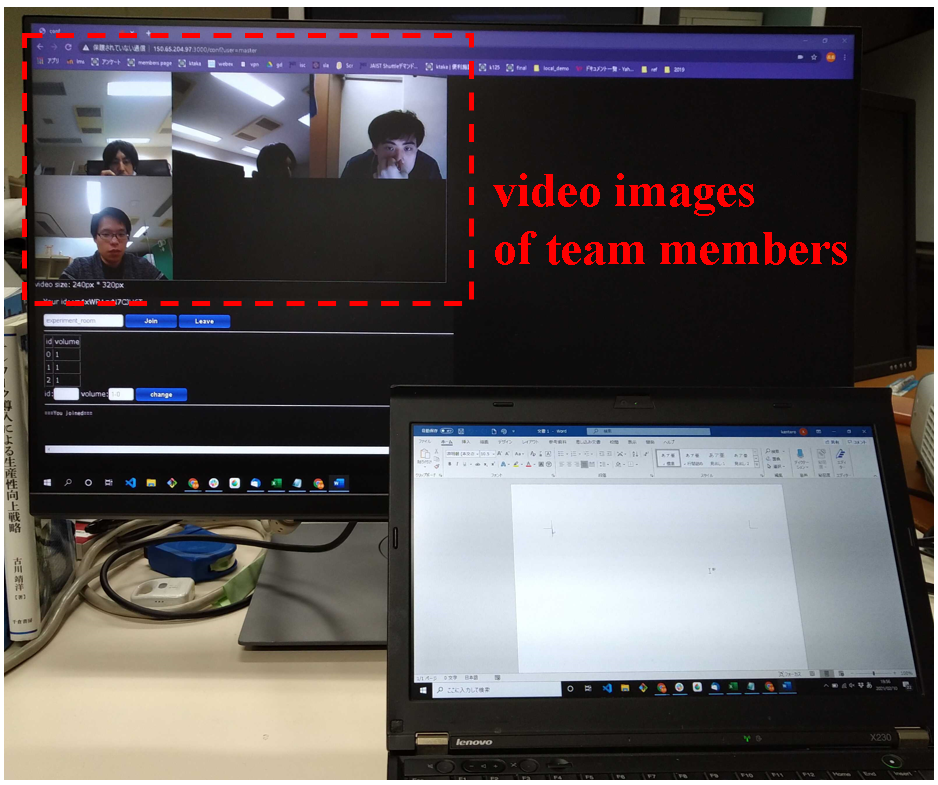}
\caption{Web chat system with ordinary video chat interface} \label{fig3u}
\end{figure}

To realize the video and audio connection among team members, we built a video chat system that provides multi-user mutual video and audio streaming connection (Fig.~\ref{fig3u}). 
 This system provides similar function as with ordinary video chat system (e.g. Zoom) but can receive commands from a gesture sensor to control the position and volume of each video.

\subsubsection{Video Projection and Calibration}
\begin{figure}[h]
\centering
\includegraphics[width=\textwidth]{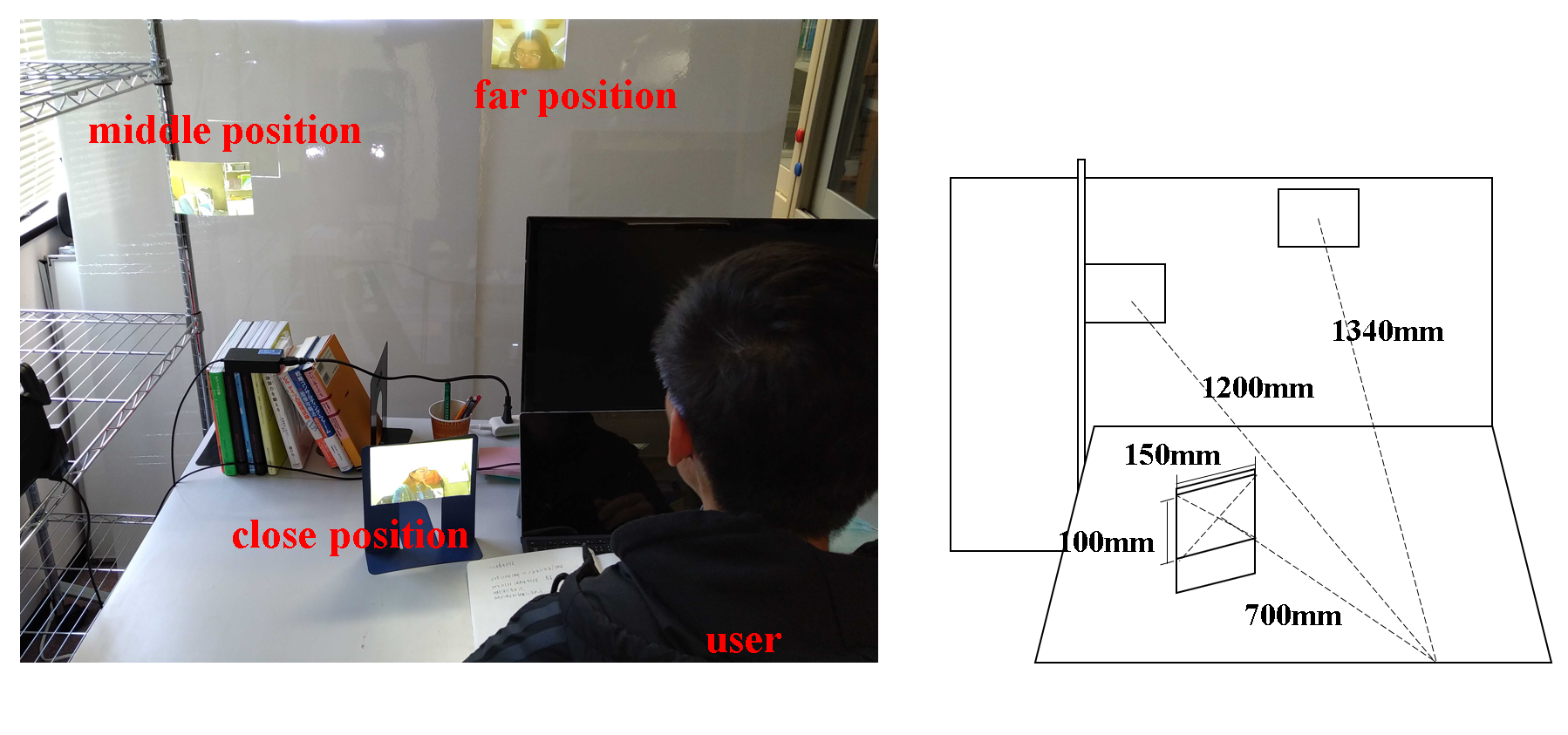}
\caption{Working scene with SpatialViewer system} \label{fig3}
\end{figure}

We determined three projection positions for team members' video images on the user's teleworking desk environment. The closest projection position is on the desk, the middle one is on a shelf, and the farthest is on the opposite wall.
 Three sticky notes were attached to the corresponding positions, and team members' videos were projected onto these notes. The work setup used with our system is shown in Fig.~\ref{fig3}.

To project video images onto a real desk environment, it is necessary to calibrate the projection space with the real space. A video image of each team member who is using the video chat system was captured and transformed to fit onto the projection planes.

\subsubsection{Gesture Control} \label{gosvp}
We developed the function of moving video images among the projected items, which also changes the corresponding audio volumes. To offer users more interactivity and freedom of manipulation, we used a gesture sensor to capture user input. Users can switch the projection locations by interacting with the sensor using hand-wave gestures.

\subsubsection{Volume Adjustment}
We developed the function of automatically adjusting the audio volume when a video's position is changed. The volume increases or decreases when a video projection moves nearer or farther. The volume of farthest video is 0.1 times that of the nearest video, and the volume of the middle-distance video is a quarter the volume of the nearest video. 

\section{Implementation of the System Prototype}
\begin{figure}[H]
\centering
\includegraphics[width=0.9\textwidth]{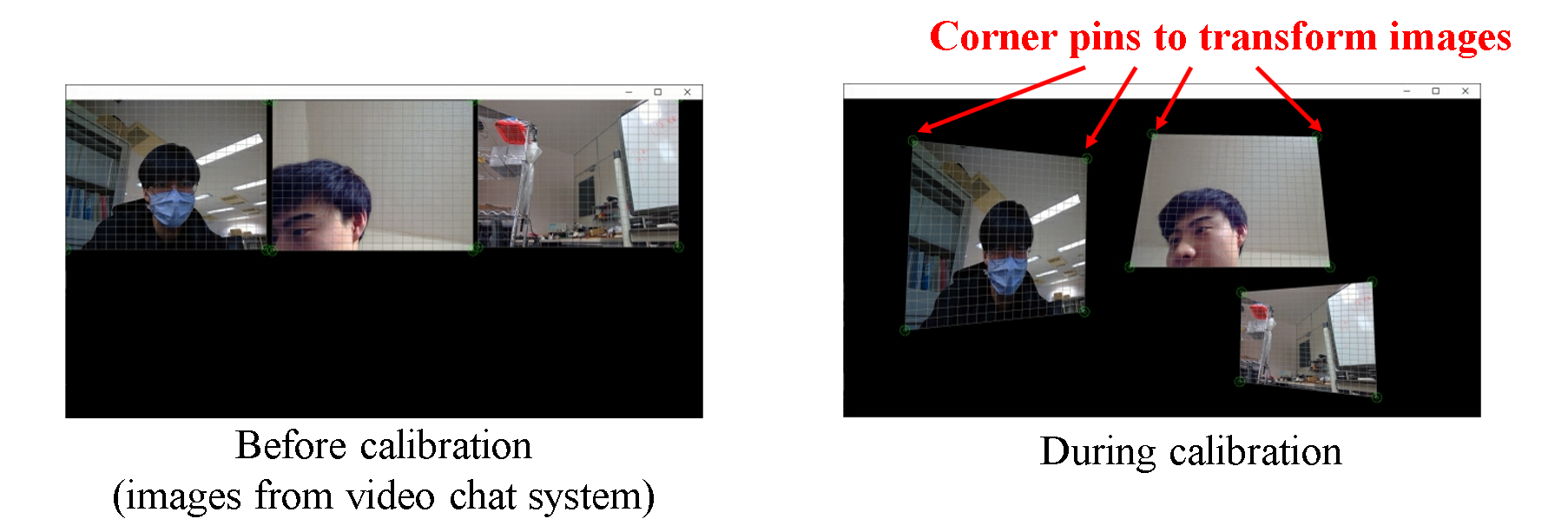}
\caption{Calibration process for multiple video frames} \label{fig4}
\end{figure}

We implemented a prototype of the proposed system. We developed a web browser-based video chat system with JavaScript and SkyWay API ~\cite{skyway}. We also developed a Processing application to transform video images for calibration and projection. We used Keystone Library in the Processing application to adjust the shape and position of the projection planes for calibration. This library enabled us to move and distort video images by dragging the pins on the corner of images 
 (see Fig.~\ref{fig4}). The Processing application and video chat system were run on a laptop PC. The audio was played by an external speaker connected to the laptop. To project the images, we places a laser 4k projector to the back and the left of the user's seat.  

To achieve gesture control, we used the Leap Motion gesture sensor. Leap Motion can send commands to the video chat system via the Processing application to switch projection positions. The WebSocket module was installed to realize a two-way real-time connection between the Processing application and the video chat system. To adjust the volume along with the projected distance, the Processing application kept recording the projection position of each video image. Each time the positions were switched, the video chat system was notified of the new position to control the audio volume. 

\section{User Study}
In this section, we first explain the experimental design and setting and then present the questionnaire design. 

\subsection{Experimental Setting}
We attempted to simulate a team collaboration teleworking environment. We recruited 11 participants for our experiment, six male and five female. Nine are master's students in their 20s attending the author's institute and two are professors in their 30s. We organized three experimental groups consisting of four participants. Each group included one professor and three students (one of the professors was included in two groups.)

Each participant took part in experiment for a total of nine hours. Each participant tried SpacialViewer (see Fig.~\ref{fig3}) alone for three hours and connected to the other members using our developed video chat interface (see Fig.~\ref{fig3u}) for six hours. All three students in each group shifted roles of SAR system user and video chat interface partner during a group of experience. 

After using the system, the students were asked to fill out the questionnaire, which asked them to evaluate the proposed system and ordinary video chat interface and record intimacy among every group members. The professors were an experimental variable to diversify group members' relationships, so they did not answer the questionnaire. 

\subsection{Questionnaire Design}
To achieve a valid evaluation, we asked several five-point Likert scale-style questions. The questions are listed below:

\begin{itemize}
\item Questions about SpatialViewer experience
    \begin{itemize}
    \item Q1: Were you interested in the situations of the projected participants?
    \item Q2: Did you want to talk to the projected participants?
    \item Q3: Did you want to switch the location of projections during the experiment?
    \item Q4: Could you feel the sense of space via the system?
    \item Q5: Did you think the system enabled you to place projections according to intimacy?
    \item Q6: Was the system useful for you compared with other teleworking applications you have used?
    \item Q7: Did you think the system provided a relaxing, comfortable teleworking environment?
    \end{itemize}
\item Questions about ordinary 
 video chat interface experience
    \begin{itemize}
    \item Q2': Did you want to talk to the participants?
    \item Q5': Did you think the system enabled you to place images according to intimacy?
    \item Q7': Did you think the video chat interface provided a comfortable teleworking environment?
    \end{itemize}
\end{itemize}

The evaluation items included the users' interaction experience (Q1 and Q2); the interactivity enabled by this system (Q3); the improvement of the sense of space (Q4); the ability to adjust projection distance according to intimacy (Q5); and the overall usefulness and comfort (Q6 and Q7) of our system. Considering compare the aspects we are most concerned about, we also asked questions analogous to Q2, Q5, and Q7 about the traditional ordinary video chat interface for comparison analysis (Q2', Q5', and Q7'). We also asked users about their sequence of intimacy (close-middle-far) to their group members and recorded the sequence of projection distance (close-middle-far) maintained the longest by participants when using the system. We then evaluated the results by analyzing the relationship between intimacy and distance.

\section{Results and Analysis}
\begin{figure}[h]
\centering
\includegraphics[width=\textwidth]{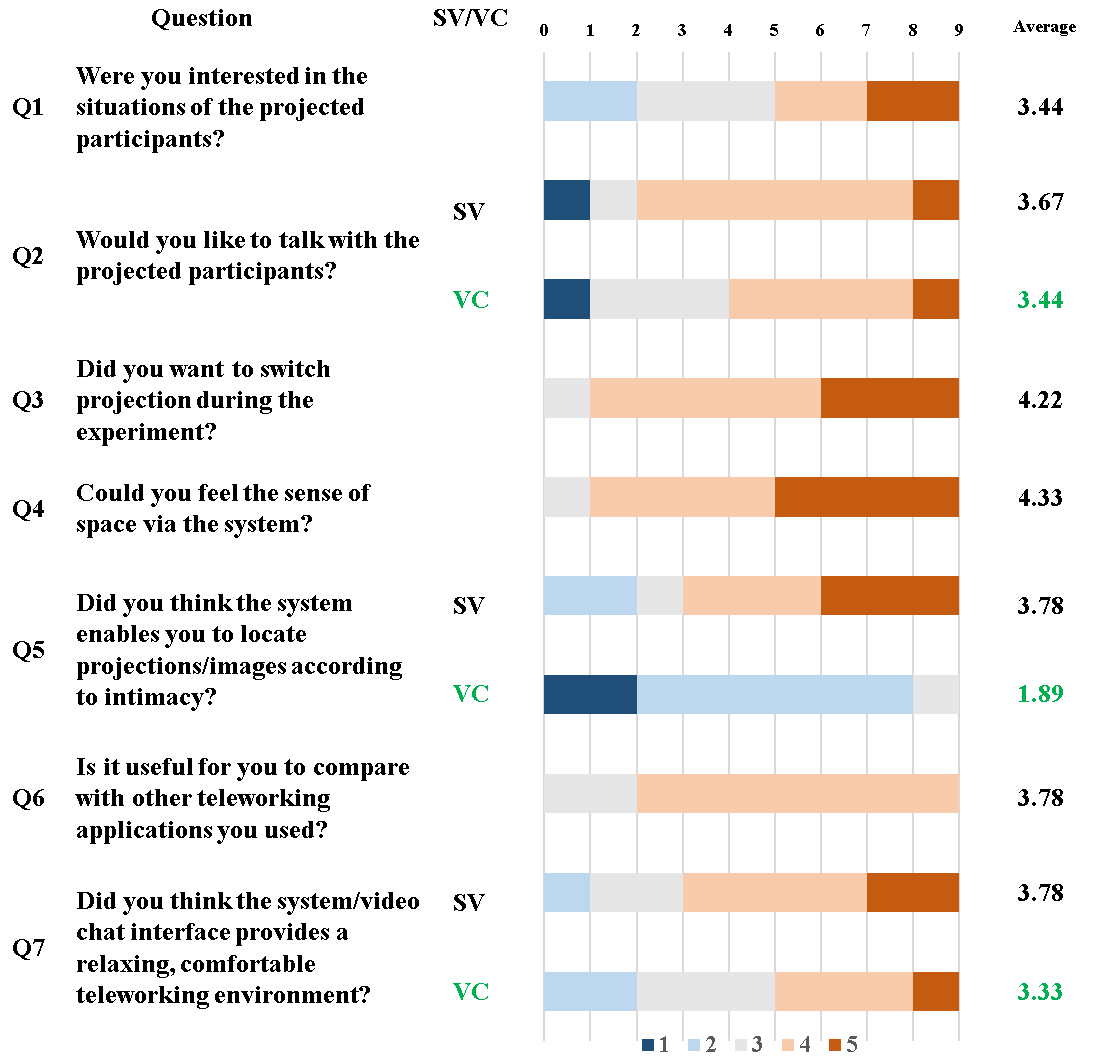}
\caption{Score results of the questionnaire questions. (SV: SpatialViewer, VC: Ordinary video chat interface)}
\label{fig5}
\end{figure}

We analyzed the results of questionnaire to verify whether our system allowed users to adjust projection distance and provided comfort.
We aggregated the scores of the questionnaire, calculated the average score of each question (see Fig.~\ref{fig5}), and conducted observation analysis of the experimental scene. Data visualization shows the positive evaluations of our system, with average scores for all questions above 3. A comparison between our projection system and a traditional video chat interface shows that our system made is easier for users to adjust distance according to intimacy (see Q5 in Fig.~\ref{fig5}) and provided a more comfortable teleworking environment (see Q7 in Fig.~\ref{fig5}). Through observing the experiment, we found that users were interested in using the interactive function provided by the gesture sensor. They used this function to adjust the projection locations to create a comfortable interaction environment.

The results also show that the individual initiative of communication did not change between SpatialViewer and ordinary video chat interface (see Q2 in Fig.~\ref{fig5}), which is consistent with the result of observation analysis. Most users conducted some daily conversations rather than talking about their work and worked silently most of the time. This might be because group members were not actually working on the same project during the experiment, which means that the need for communication was insufficient. Although the participants did not communicate much, they tended to place the projection of their communication target closer to them during the conversation, which shows that the system realized the spatial sense of communication.

Participants could freely control interpersonal distances during the experiment. Table\ref{tab1} shows a matrix of the relationship between projection distance and degree of intimacy of users. Seven of the nine experimental results conformed to the ideal matrix distribution (diagonal distribution), which indicates that the projection distance was positively correlated with intimacy. For instance, many users chose to put the professor farther away, and most users had the most distant relationship with the professor. This proves that, even when teleworking, interpersonal distance is affected by intimacy.

However, some results did not conform to expectations. We interviewed each participant and received some valuable feedback. For instance, they told us that they cared more about how the member's appearance in the video made them feel than intimacy. They preferred to place the image of a group member with good posture and a moderate distance from the camera closest to themselves. If another member was too close to the camera or put the camera in the wrong position, this could cause stress and discomfort. The interviews indicated that there is no single factor affecting the projection locating preference of users; intimacy and visual comfort may be other influencing factors.

\begin{table}
\centering
\caption{Cross tabulation matrix of projection distance and intimacy of group members.}\label{tab1}
\begin{tabular}{|l|l|l|l|}
\hline
\diagbox{Dist.}{Intim.} & close & middle & far\\
\hline
close & 7 & 1 & 1\\
middle & 0 & 8 & 1\\
farther & 2 & 0 & 7\\
\hline
\end{tabular}
\end{table}

In the interviews, many participants reported that our system provided a novel way of teleworking, and they felt it would be valuable to them (see also Q6 in Fig.~\ref{fig5}). A convenient use of the system was the ability to move the projections of members who they were talking to. Another advantage was the ability to adjust the projection distance, in accordance with our original expectations.

We also asked users about any concerns with or desired improvements for the system. Most of the participants' recommendations focused on interaction and communication functions. Three participants wished the interactive function could adjust the size of projections. Two participants wanted a device promoting willingness to communicate, 
 and one participant wanted a method of private communication to talk with specific group members without being heard by the others. These advises show that our system is weak in motivating interaction and communication.

\section{Conclusion}
In this work, we proposed a remote work sharing system that takes into consideration the levels of intimacy among workers. Users can customize the location of the projected videos and audio volume corresponding to each team member. We conducted an experiment to evaluate the system. The questionnaire results show that the system allowed users to adjust projection distances according to intimacy, provided a sense of space, and thus improved users' comfort with remote work compared with a traditional video chat interface.

We developed a prototype system for use in experiments and laboratory demonstrations. According to the feedback received from participants, the system needs to be improved in terms of interactivity and convenience. Also, we will improve the user interface and append more functions like controlling projection position at any place in user's desk environment 
 in the future. The experiment setting will also be reconsidered. The system should be evaluated under actual working conditions and compared with other applications proposed in previous studies.


\bibliographystyle{splncs03_unsrt}
\bibliography{ref}
\end{document}